\documentclass[prb,floatfix,twocolumn,showpacs,amsmath,amssymb]{revtex4}
\usepackage{graphicx}
\usepackage{dcolumn}
\usepackage{bm}

\newcommand{\beq}{\begin{equation}}
\newcommand{\eeq}{\end{equation}}
\newcommand{\beqa}{\begin{eqnarray}}
\newcommand{\eeqa}{\end{eqnarray}}
\newcommand{\kvec}{{\bf k}}

\newcommand{\Kvec}{{\bf K}}
\newcommand{\rvec}{{\bf r}}
\newcommand{\Rvec}{{\bf R}}

\begin{document}

\title{Tunnelling matrix elements with antiferromagnetic Gutzwiller wave functions}

\author{Andrea Di Ciolo, Luca F. Tocchio, Claudius Gros}
 \affiliation{Institute for Theoretical Physics, Goethe-University
Frankfurt, Max-von-Laue Stra{\ss}e 1, D-60438 Frankfurt am Main, Germany.}
 

\date{\today} 

\begin{abstract}

We use a generalized Gutzwiller Approximation (GA) elaborated to evaluate matrix elements with partially projected wave functions and formerly applied to homogeneous systems.\cite{Fuk05, Ede07}
In the present paper we consider projected single-particle (hole) excitations for electronic systems with antiferromagnetic (AFM) order and obtain the corresponding tunnelling probabilities. The accuracy and the reliability of our analytical approximation is tested using the Variational Monte Carlo (VMC). Possible comparisons with experimental results are also discussed. 

\end{abstract}


\maketitle

\section{Introduction}

Strongly correlated electronic systems can be studied by means of several methods like quantum Monte Carlo\cite{Fou01,Cape06,Bec09}, dynamical mean-field theory\cite{Geo96},  numerical and functional renormalization group\cite{Ber02, Scho05, Kop09}, density functional theory\cite{Ani93, Czy94, Ylv09}, Gutzwiller method\cite{Gut63} and slave bosons.\cite{Kot86}   
 Most approaches for the strong coupling regime are numerical: they can explore cases difficult to be treated analytically but they are typically computer-time consuming and not always allow a full physical insight into the investigated quantities. On the other side, analytical computations lead to results  in form of explicit relations but they often require some approximations or ansatz not totally under control in their accuracy. This is also the case of the variational studies which use the Guztwiller wave functions and evaluate the observables via the Gutzwiller approximation. We focus our paper on the evaluations of matrix elements with Gutzwiller functions, motivated by the past success of studies with these projected wave functions. In fact, they provided significant progresses in the characterization of strongly correlated systems\cite{Vol84}, also to study the effective models proposed to explain the properties of high-temperature superconductors.\cite{And87, And87-II, Bas87} We also intend to establish more precisely the reliability and the drawbacks of our approximated analytical computations: to this aim, we  calculate some test quantities also with the VMC, which provides exact results within the statistical uncertainties.  

Gutzwiller proposed to study electronic systems with repulsive onsite interactions using projected wave functions of the form $|\Psi\rangle = P|\Psi_0\rangle$, where $P$ is a projector which suppresses the double occupancies of the uncorrelated trial state $|\Psi_0\rangle$.\cite{Gut63} If the latter is the homogeneous state, the projector is $P = \prod_{i}(1-n_{i\uparrow} n_{i\downarrow})$ where $n_{i\sigma}=c^\dag_{i\sigma}c_{i\sigma}$, with $c^\dag_{i\sigma}$ ($c_{i\sigma}$) the creation (annihilation) operator for an electron of spin $\sigma$ on a site $i$. 

In the GA, one evaluates the expectation values of operators $\hat{O}$ using the projected wave functions  $\Psi$. The effects of the projection on the state $|\Psi_0 \rangle$ are approximated by a classical statistical weight factor $g_o$ which multiplies the unprojected quantum result:\cite{Gut63,Vol84}
\begin{equation}
\frac{\langle \Psi| \hat{O} |\Psi \rangle}{\langle \Psi |\Psi \rangle }\ =\ g_o\
\frac{\langle \Psi_0| \hat{O} |\Psi_0\rangle}{\langle \Psi_0|\Psi_0\rangle} \ .
\end{equation}
Basically, the projection operator $P$ reduces the number of the allowed states in the Hilbert space, and in the simplest approximation this renormalization is taken into account through combinatorial factors. In order to obtain the renormalization factors, one counts the number of states that contribute to $\langle \Psi| \hat{O} |\Psi \rangle$ and to $\langle \Psi_0| \hat{O} |\Psi_0\rangle$, and their ratio provides the factor. If for example $\hat{O}$ is the kinetic operator $\hat{T}_{ij\sigma} = c^\dagger_{i\sigma} c^{}_{j\sigma} +  c^\dagger_{j\sigma} c^{}_{i\sigma}$, the corresponding factor $g_{t\sigma}$ expresses the reduction of the kinetic energy for correlated electrons. Such renormalization factors are obtained for example in Ref.~\onlinecite{Gut63, Vol84, KoNa07, Fuk08, Zha88,Geb90,Sei01} and also in equivalent slave-boson approaches.\cite{Kot86,Lav90} In our present work, we use Gutzwiller factors obtained neglecting any noncombinatorial configuration dependences and intersite correlations. For further discussions, the interested reader can consider Refs.~\onlinecite{KoNa07,Fuk08}, for example.
  As we will explain better in Sect.~II, we also introduce an auxiliary Gutzwiller projector $P'$ to handle computations with partially projected wave functions: $P'$ suppresses the double occupancies for all the lattice sites but for a reservoir site.\cite{Fuk05, Ede07} 

In this paper we consider projected single-particle (hole) excitations and evaluate the corresponding tunnelling probabilities for an AFM, {\em i.e.} a system on a bipartite lattice with collinear spin order and staggered magnetization $m$.  
We consider the AFM case, also because this is a recurring wave function for the ground state of (lowly-doped) strongly correlated systems.

We emphasize that in our approach the results are due only to the physical properties of the trial state and not to the choice of a specifical Hamiltonian: in this sense, they are model-independent but they are not universal, because they rely on the features of the chosen Gutzwiller wave function (Fermi sea, BCS superconductor, AFM...)

Our results can be helpful to understand the spectra probed in one-particle spectroscopy experiments (tunnelling, photoemission) for materials displaying a long-range collinear AFM order. Actually, to date there are not many tunnelling experiments dedicated for strongly correlated materials in their long-range AFM region. In fact, the standard Scansion Tunnelling Microscopy (STM) method can not detect spin structures\cite{Fis07} and the improved Spin-Polarized STM can not be applied for insulating materials.\cite{Wie09, Koh07} In order to study the AFM order in correlated insulators, 
a new spin-probe method has been developed: the Magnetic Exchange Force Microscopy\cite{Wie09,Kai07}, which does not involve one-particle processes.  
We will discuss possible future comparisons with experiments more extensively in Sect.~VI. 

This paper is organized as follows: in Sect.~II we illustrate the method and the formalism; in Sect.~III we provide the normalization for the projected particle and hole excitations and in Sec.~IV the probability for the particle and hole tunnelling; in Sect.~V we check the accuracy of the GA using VMC computations. In Sect.~VI we discuss the physical meaning of our findings and then we come to the Conclusions. In Appendix 1 we illustrate more specifically how to obtain the GA hopping factors introduced in Sect.~II and present the full derivation of the analytical results of Sect.~III and ~IV; in Appendix 2 we present some additional VMC results for projected BCS superconductors, in order to understand better some issues discussed in Ref.~\onlinecite{Fuk05}. 

\section{Projected wave functions}\label{sec:formalism}
In this Section we will recall the crucial steps of the generalized scheme presented in Refs.~\onlinecite{Fuk05, Ede07} and show how to adapt the formalism for the AFM wave functions.

We consider an electronic system on a $L$-site lattice, where the position of each atom is given by $\rvec_{i\alpha} = \Rvec_i + \delta_\alpha$: $\Rvec_i$ belongs to the Bravais lattice and $\delta_\alpha$ fixes the atomic site in the unit cell and then determines its sublattice. We will often use the shorthand notation $\rvec_{i\alpha}= i+\alpha$ and set the lattice parameter $a$=1.

Considering states with a single particle (hole) excitation with respect to the ground state, it is convenient to work with the following partially projected wave function,
\begin{eqnarray}
 |\Psi_{l\gamma}'\rangle \ &=&\ P_{l\gamma}'|\Psi_0\rangle.
\label{psi_l}
\end{eqnarray}
Double occupancy is projected out for all the sites, but for the reservoir site $l\gamma$.  
 In order to study AFM wave functions, the projector in Eq.~(\ref{psi_l}) has the form 
\begin{eqnarray}
P_{l\gamma}' =  \prod_{i\alpha\ne l\gamma}(1-n_{i\alpha\uparrow} n_{i\alpha\downarrow}),
\label{Pl}
\end{eqnarray} 
whereas the parent fully projected state $|\Psi\rangle$ is 
\begin{eqnarray}
|\Psi\rangle = P|\Psi_0\rangle,
\label{psi}
\end{eqnarray}
with the projector 
\begin{eqnarray}
P =  \prod_{i\alpha}(1-n_{i\alpha\uparrow} n_{i\alpha\downarrow}). 
\label{Ptot}
\end{eqnarray} where $n_{i\alpha\sigma}$ is the projected spin-$\sigma$ electronic density for a site $i$ on the sublattice $\alpha$. The lattice is bipartite and divided into two sublattices, A and B: we fix as sublattice A the one with $n_{\alpha\uparrow} > n_{\alpha\downarrow}$.

In order to work in the momentum space, we adopt the following convention to perform the Fourier Transform on a poliatomic unit cell:
\begin{eqnarray*}
c^{\dag}_{i\alpha\sigma} &=& \dfrac{1}{\sqrt{L}} \sum_{\kvec\Kvec} e^{i\kvec\cdot\Rvec_i} e^{i(\kvec+\Kvec)\cdot \delta_\alpha}c^{\dag}_{\kvec+\Kvec,\sigma},
\label{ckK}
\end{eqnarray*}
with $\kvec$ belonging to the Reduced Brillouin Zone and $\Kvec$ to the reciprocal lattice.
We can use a canonical transformation $\phi$ to introduce the quasiparticle operators $\alpha_{\kvec}$ from the original fermionic $c_{\kvec}$ operators:
\begin{eqnarray}
\alpha^{\dag}_{\kvec\lambda\sigma} &=& \sum_{\Kvec} \phi^*_{\lambda\Kvec}(\kvec\sigma) c^{\dag}_{\kvec+\Kvec,\sigma}. 
\label{Transf}
\end{eqnarray}
Then we can present Eq.~(\ref{Transf}) in an explicit form, using this Bogolubov transformation for the AFM:
\begin{eqnarray}
\alpha^{\dag}_{\kvec 1\sigma} &=& u_{\kvec}c^{\dag}_{\kvec\sigma} + \sigma v_{\kvec}c^{\dag}_{\kvec+\Kvec_{AF},\sigma} \nonumber \\ 
\alpha^{\dag}_{\kvec 2\sigma} &=& -\sigma v_{\kvec}c^{\dag}_{\kvec\sigma} + u_{\kvec}c^{\dag}_{\kvec+\Kvec_{AF},\sigma}. 
\label{Bog}
\end{eqnarray}  
being $\Kvec_{AF}$ the AFM ordering wave vector.                      
We consider the unprojected state $|\Psi_0\rangle$ which displays long-range AFM order, as found in previous studies:\cite{Zha88,Yok87-1, Yok87-2, Gro89} 
\begin{eqnarray}
|\Psi_0\rangle &=& \prod_{\kvec\sigma} \alpha^{\dag}_{\kvec1\sigma}|0\rangle \nonumber \\ &=& \prod_{\kvec\sigma}[u_{\kvec}c^{\dag}_{\kvec\sigma} + \sigma v_{\kvec}c^{\dag}_{\kvec+\Kvec_{AF},\sigma}]|0\rangle;  
\label{AFPsi0}
\end{eqnarray}
$|0\rangle$ is the electron vacuum and \begin{eqnarray}
u^2_\kvec &=& \dfrac{1}{2}\Big(1-\dfrac{\epsilon^0_{\kvec\sigma}}{\sqrt{(\epsilon^0_{\kvec\sigma})^2+ \Delta^2}}\Big) \nonumber \\ 
v^2_\kvec &=& \dfrac{1}{2}\Big(1+\dfrac{\epsilon^0_{\kvec\sigma}}{\sqrt{(\epsilon^0_{\kvec\sigma})^2+ \Delta^2}}\Big),
\label{uvk}
\end{eqnarray} 
where $\epsilon_{\kvec\sigma}^0$ is the bare electronic energy spectrum and $\Delta$ the staggering field. 
The condition $u^2_{\kvec} + v^2_{\kvec} = 1$ comes from the normalization of the wave function $\Psi_0$. 

\subsection*{Gutzwiller approximation}\label{subsec:AFM}

The general form of the hopping renormalization factors $g_{t\alpha\beta\sigma}$ for a SDW (Spin Density Wave) is 
\begin{eqnarray}
g_{t\alpha\beta\sigma} &=& \sqrt{\dfrac{1}{1-\dfrac{2n^0_{i\alpha\sigma} n^0_{i\alpha,-\sigma}}{n}}\dfrac{1}{1-\dfrac{2n^0_{j\beta\sigma} n^0_{j\beta,-\sigma}}{n}}} \cdot \nonumber \\ &\cdot& \dfrac{(1-n)[(1-n^0_{i\alpha,-\sigma})(1-n^0_{j\beta,-\sigma})]^{1/2}}{[(1-n^0_{j\beta\sigma})(1-n^0_{i\alpha\sigma})]^{1/2}}
\label{gSDW}
\end{eqnarray}
where $n^0_{i\alpha\sigma}$ ($n^0_{j\beta\sigma}$) is the unprojected spin-$\sigma$ density for a site $i$ ($j$) on the sublattice $\alpha$ ($\beta$) and $n$ the charge density (the filling). 
Eq. (\ref{gSDW}) can be obtained from the general expression of the Gutzwiller factor $g_{t}$, as we derive in Appendix~1A.  
If we consider the case of sites $i\alpha$ and $j\beta$ on different sublattices in an AFM, we obtain the factor
\begin{eqnarray*}
g_{tAB} = \frac{1-n}{1-2 n^0_\uparrow n^0_\downarrow/n} 
\label{gtAB}
\end{eqnarray*}
which gives the renormalization for the nearest-neighbour hopping ($g_{tAB}\equiv g_{tAB\sigma}$; $n^0_\uparrow$ and $n^0_\downarrow$ are taken on the same sublattice).

In order to calculate the excitation norms and the tunnelling matrix elements, we work with expectation values such as
\begin{equation}
\frac{\langle \Psi_{l\gamma}'| \hat{O} |\Psi_{l\gamma}'\rangle}{\langle \Psi_{l\gamma}'|\Psi_{l\gamma}'\rangle }\ =\ g_\gamma'\
\frac{\langle \Psi_0| \hat{O} |\Psi_0\rangle}{\langle \Psi_0|\Psi_0\rangle } 
\label{renorm_l}
\end{equation}
that generalize the Gutzwiller renormalization scheme to partially projected wave functions.

Thus we need the normalization $\langle \Psi_{l\gamma}'|\Psi_{l\gamma}'\rangle$ to evaluate the parameters $g_\gamma'$ in Eq.~(\ref{renorm_l}). To this aim, we introduce the quantity: 
\begin{eqnarray}
 X_{\gamma}\ &=& \frac{\langle \Psi |\Psi \rangle}{\langle \Psi_{l\gamma}'|\Psi_{l\gamma}'\rangle} = \frac{\langle \Psi_0|P P |\Psi_0\rangle}{\langle \Psi_0|P_{l\gamma}' P_{l\gamma}'
  |\Psi_0\rangle }\  \label{def_X} \\ 
            &=& \dfrac{{\mathcal C}_{d_{l\gamma}=0}}{{\mathcal C}_{d_{l\gamma}=0} + {\mathcal C}_{d_{l\gamma}= 1}} \nonumber
\end{eqnarray}
where ${\mathcal C}_{d_{l\gamma}=0}$ is the number of states with empty or singly occupied reservoir site and ${\mathcal C}_{d_{l\gamma}= 1}$ the ones with doubly occupied reservoir site. Then, $X_{\gamma}$ is the ratio between the norm of the fully projected state and the norm of the state with one reservoir site.
For the AFM the norms $\langle \Psi |\Psi \rangle$ and $\langle \Psi_{l\gamma}'|\Psi_{l\gamma}'\rangle$ take the following form:  
\begin{widetext}
\begin{eqnarray}
\langle \Psi |\Psi \rangle = \sum_{N_{\alpha\uparrow},N_{\alpha\downarrow}} &&\binom {L/2}{N_{\alpha\uparrow},N_{\alpha\downarrow}, N_{h\alpha}} \binom {L/2}{N_{\uparrow}-N_{\alpha\uparrow},N_{\downarrow}-N_{\alpha\downarrow}, L/2 - N + N_\alpha} w_\alpha\nonumber  \\ 
\langle \Psi_{l \gamma}'|\Psi_{l \gamma}'\rangle = \sum_{N_{\alpha\uparrow},N_{\alpha\downarrow}} && \left [\binom {L/2}{N_{\alpha\uparrow},N_{\alpha\downarrow}, N_{h\alpha}} \binom {L/2}{N_{\uparrow}-N_{\alpha\uparrow},N_{\downarrow}-N_{\alpha\downarrow}, L/2 - N + N_\alpha}w_\alpha  \nonumber \right . \\  &+&  \left . \binom {L/2}{N_{\alpha\uparrow},N_{\alpha\downarrow}, N_{h\alpha}} \binom {L/2-1}{N_{\uparrow}-N_{\alpha\uparrow}-1,N_{\downarrow}-N_{\alpha\downarrow}-1, L/2 - N + N_\alpha+1}w_\alpha \phantom{\Big)}\right] \nonumber \\ 
\label{XAFM}
\end{eqnarray}
\end{widetext}
where $w_\alpha = e^{\tilde{\beta} m(m-m_\alpha)}$, being $\tilde{\beta}$ a constant. $w_\alpha$ is a statistical weight associated to any electronic configuration: it is 1 for the homogeneous case and in first approximation for the AFM in case of moderate magnetizations. 

The $L$-site lattice with $N$ electrons ($N_\sigma$ per spin $\sigma$) is bipartite in two sublattices with $L/2$ sites. We indicate with $N_\alpha$ the number of electrons for each sublattice ($N_{\alpha\sigma}$ per spin $\sigma$) and with $N_{h\alpha}$ the corresponding number of holes, being $N_{h\alpha} = L/2 - \sum_{\sigma}N_{\alpha\sigma}$. For a given configuration of the electrons on the two sublattices the total magnetization is $M_\alpha = N_{\alpha\uparrow} - N_{\alpha\downarrow} = (L/2)m_\alpha$ with $m_\alpha$ the correspondent magnetization density for that configuration. Clearly, the possible values of $M_\alpha$ are in general different from the thermodynamical total magnetization $M = (L/2)m$.
In Eq.~(\ref{XAFM}) the sums have the constraints $M \leq N_{\alpha\uparrow} \leq N_\uparrow$; $0 \leq N_{\alpha\downarrow} \leq N_{\downarrow}-M$, with $N_\sigma = N/2$. 

For the homogeneus state ($m$=0), our Eq.~(\ref{XAFM}) recovers Eq.~(10) of Ref.~\onlinecite{Fuk05} in the thermodynamical limit. For the N\'eel state ($m=n$), all the spins $\uparrow$ are on the same sublattice and all the spins $\downarrow$ on the other one: this implies that the double occupancy in the reservoir site is not possible and then $X_{\gamma}=1$: in this case, the generalized GA reduces to the standard GA and the auxiliary projector $P'_{l\gamma}$ to the projector $P$. 

Eq.~(\ref{XAFM}) cannot be presented in a simple form, neither in the thermodynamical limit, and thus we check if a formula guessed after the homogeneous result can represent a reasonable approximation. The candidate form is
\begin{figure}[tp]
\centering
\includegraphics[width=8.65cm]{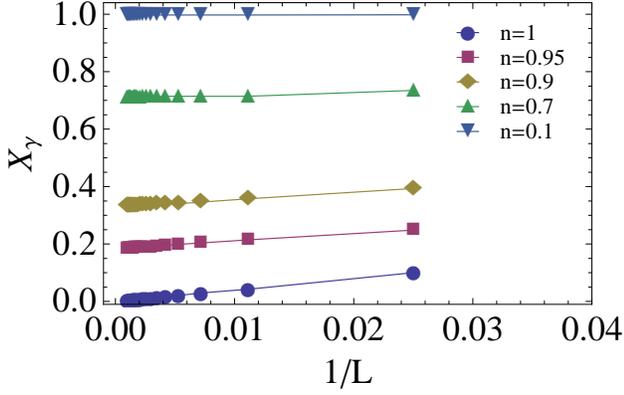} 
\caption{Factor $X_{\gamma}$ as a function of $1/L$, for the magnetization $m=n/2$ and several fillings $n$ ($L$ is the number of the lattice sites).}
\label{XL}
\end{figure} 
\begin{eqnarray}
X^{ap}_{\gamma} &=&  \dfrac{1-n}{(1-n_{l\gamma\uparrow})(1-n_{l\gamma\downarrow})} = X 
\label{xguess}
\end{eqnarray}
which fulfils the known behaviour of $X_{\gamma}$ both for $m$=0 and $m=n$. Thus we investigate if Eq.~(\ref{xguess}) approximates Eq.~(\ref{XAFM}) adequately for intermediate $m$. 

At first, in Fig.~\ref{XL} we perform an analysis of the finite-size effects affecting $X_{\gamma}$ computed using Eq.~(\ref{XAFM}), for several fillings and for several lattices sizes (from $L=40$ up to $L \sim 10^3$). Clearly the deviation of the obtained values with respect to the average ones reduces more and more upon increasing $L$ and we conclude that $X_{\gamma}$ computed on $L\sim 10^3$ site-lattices is representative of the result for $L\rightarrow\infty$ and will be compared with $X^{ap}_{\gamma}$ evaluated using Eq.~(\ref{xguess}).  

\begin{figure}[tp]
\centering
\includegraphics[width=8.65 cm]{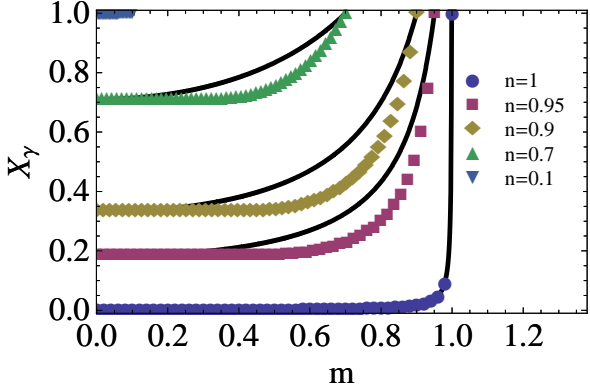} 
\caption{Comparison of the factors $X_{\gamma}$ (in symbols) and $X^{ap}_{\gamma}$ (in solid lines) as a function of the magnetization $m$, for several fillings $n$ on a 1000-site lattice.}
\label{Xm}
\end{figure}

\begin{figure}[tp]
\centering
\includegraphics[width=8.65 cm]{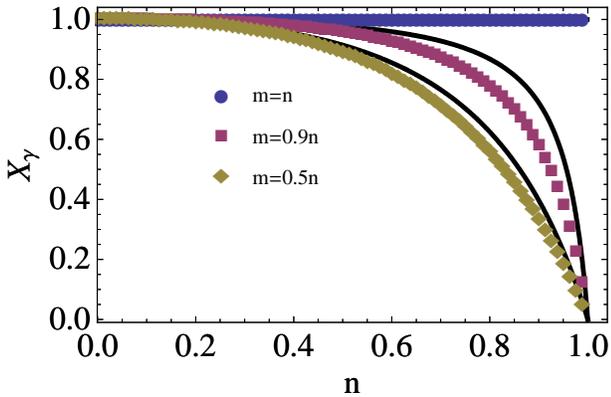}
\caption{Comparison of the factors $X_{\gamma}$ (in symbols) and $X^{ap}_{\gamma}$ (in solid lines) as a function of the filling $n$, for three magnetizations $m$ on a 4000-site lattice.}
\label{Xn}
\end{figure}

Then, in Figs.~\ref{Xm} and ~\ref{Xn}, we compare systematically the behaviour of $X_{\gamma}$ and $X^{ap}_{\gamma}$. In Fig.~\ref{Xm} we consider the trends of $X_\gamma$ as a function of $m$ for several fillings and in Fig.~\ref{Xn} as a function of $n$ for several magnetizations. The general outcome is that $X^{ap}_{\gamma}$ is a very good approximation of $X_{\gamma}$ for small $m$ and good for large $m$. In fact, the approximation is quite fair, because even for intermediate $m$ (where the accord is less satisfactory), the quantitative discrepancy with the exact results is moderate (about up to 20$\%$) and the qualitative trend is similar. 

We conclude that Eq.~(\ref{xguess}) is a controlled approximation of Eq.~(\ref{XAFM}) for large $L$ and therefore we decide to adopt Eq.~(\ref{xguess}) for $X_{\gamma}$ in the analytical derivations of the following Sections.

\subsection*{Exact relations in the GA}
For the occupancy of the reservoir site $l\gamma$, the following exact expressions hold
\begin{eqnarray} 
 \langle (1-n_{l\gamma\uparrow})(1-n_{l\gamma\downarrow})\rangle _{\Psi_{l\gamma}'} & =& X_{\gamma}(1-n_{l\gamma}) \nonumber \\
\label{psi_l_single}
\langle n_{l\gamma\sigma}(1-n_{l\gamma,-\sigma})\rangle _{\Psi_{l\gamma}'} & =& X_{\gamma} n_{l\gamma\sigma} \nonumber \\
\label{psi_l_double}
\langle d_{l\gamma}\rangle _{\Psi_{l\gamma}'} \ \equiv\  
 \langle n_{l\gamma\uparrow}n_{l\gamma\downarrow}\rangle _{\Psi_{l\gamma}'} & =& 1-X_{\gamma}
\label{psi_l_empty}
\end{eqnarray}
where $\langle ... \rangle _{\Psi_{l\gamma}'}\equiv
\langle \Psi_{l\gamma}'|  ... |\Psi_{l\gamma}' \rangle / \langle \Psi_{l\gamma}'|\Psi_{l\gamma}'\rangle$.\cite{Fuk05}

In the following Sections we will use $d_\gamma$ as shorthand notation for $\langle d_{l\gamma}\rangle _{\Psi_{l\gamma}'} = d_{l\gamma}$.

\section{Single-particle excitations of the projected antiferromagnet}\label{sec:excit}
We consider the particle excitation
\begin{equation}
|\Psi_{k+K,\sigma}^+\rangle  \,=\, P c_{\kvec+\Kvec,\sigma}^\dagger|\Psi_0\rangle,
\label{psi+}
\end{equation}
and the hole excitation
\begin{equation}
|\Psi_{k+K,\sigma}^-\rangle  \,=\, P c_{\kvec+\Kvec,\sigma}|\Psi_0\rangle.
\label{psi-}
\end{equation}
For the computations with $|\Psi_{k+K,\sigma}^\pm\rangle $, we need to determine the norms $N_{k+K,\sigma}^\pm=\langle \Psi_{k+K,\sigma}^\pm |\Psi_{k+K,\sigma}^\pm\rangle$. 
We will present here the results of our computations for the investigated matrix elements where we used the GA; the full analytical derivation can be found in Appendix~1B (particle excitations) and 1C (hole excitations). 


\subsection{Particle excitation}\label{subsec:partexcit}

We intend to evaluate the norm of the projected particle excitations:
\begin{eqnarray} 
N_{k+K,\sigma}^+ \ &=&  
\langle \Psi_0|c_{\kvec+\Kvec,\sigma}^{\phantom{\dagger}}P Pc_{\kvec+\Kvec,\sigma}^\dagger|\Psi_0\rangle.
\label{partnorm}
\end{eqnarray}
After the due computations in the GA, Eq.~(\ref{partnorm}) takes this form:
\begin{eqnarray}
\dfrac{N_{k+K,\sigma}^+}{N_G} &=& (1-n) + \dfrac{1}{2}(g_{tA\sigma}n^0_{A\sigma}+g_{tA,-\sigma}n^0_{A,-\sigma}) \nonumber \\ &-& g_{tAB}\langle c^{\dagger}_{\kvec+\Kvec,\sigma}c^{\phantom{\dagger}}_{\kvec+\Kvec,\sigma}\rangle_{\Psi_0},  
\label{norm_psiAFM+}
\end{eqnarray}
where $N_G=\langle \Psi | \Psi \rangle$, 
$g_{tA\sigma} \equiv g_{tAA\sigma}$ [see Eq.~(\ref{gSDW})] and $n^0_{A\sigma}$ is the unprojected spin-$\sigma$ density for the $A$-sublattice. 
Thanks to Eq.~(\ref{Bog}), we can present Eq.~(\ref{norm_psiAFM+}) in the following way: 
\begin{eqnarray}
\dfrac{N_{k\sigma}^+}{N_G} &=& (1-n) + \dfrac{1}{2}(g_{tA\sigma}n^0_{A\sigma} + g_{tA,-\sigma}n^0_{A,-\sigma}) \nonumber \\ 
&-& g_{tAB}(u^2_{\kvec}n^{0}_{\kvec1\sigma} + v^2_{\kvec}n^{0}_{\kvec2\sigma}) \nonumber \\
\dfrac{N_{k+K_{AF},\sigma}^+}{N_G} &=& (1-n) + \dfrac{1}{2}(g_{tA,-\sigma}n^0_{A,-\sigma} + g_{tA\sigma}n^0_{A\sigma}) \nonumber \\ 
&-& g_{tAB}(v^2_{\kvec}n^{0}_{\kvec 1\sigma} + u^2_{\kvec}n^{0}_{\kvec 2\sigma})
\label{N++}  
\end{eqnarray}  
\begin{center}
\begin{figure}[tp]
\includegraphics[width=8.65cm]{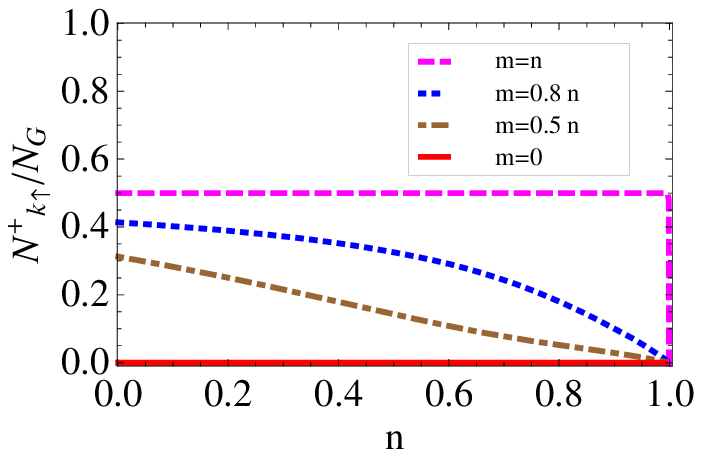}
\caption{Ratio between the particle excitation norm $N^+_{\kvec\uparrow}$ and the norm $N_G$ of the correlated state $\Psi$,  as a function of the filling $n$, for several magnetizations $m$.}
\label{Npn}
\end{figure}
\end{center}
where $n^{0}_{\kvec\lambda\sigma} = \langle\alpha^\dag_{\kvec\lambda\sigma}\alpha^{\phantom{}}_{\kvec\lambda\sigma}\rangle_{\Psi_0}$, {\em i.e.} the unprojected momentum distribution function associated to the magnetic quasiparticle state $|\Psi^0_{k\lambda\sigma}\rangle = \alpha^\dagger_{\kvec\lambda\sigma}|0\rangle$.

For example, in Fig.~\ref{Npn} we report the behaviour of the ratio between the particle excitation norm $N_{k\uparrow}^+$ and the norm $N_G$ of the correlated state $\Psi$, as a function of the electronic filling $n$ for some relative magnetizations $m/n$. This study helps us determine for which parameters the addition of one particle to the ground state is more or less favoured or even forbidden. Let us consider the case of an electron with $k<k_F$ for $m=0$, being $k_F$ the Fermi momentum (we will consider the same case for Figs.~\ref{Nhn} and \ref{Mhn} as well). In such paramagnetic case, the particle excitation is not allowed for any fillings. Upon increasing the ratio $m/n$, this scenario is removed and particle excitations are possible for any finite doping. We can notice a typical trend associated to the increase of $m/n$: the probability is maximal for dilute systems and then decreases monotonically for larger fillings. The extreme situation occurs for the N\'eel state: the ratio $N_{k\uparrow}^+ / N_G$  is constantly 1/2 until the doping closest to zero; for $n=1$ the probability for the particle excitation drops to zero and this is an isolated point.

In general, the size of the magnetization affects a lot the entity of this excitation norm, but one feature persists for any value of $m/n$: it is forbbiden to have a projected particle-excited state at half-filling.     

\begin{widetext}
\subsection{Hole excitation}\label{subsec:holexcit}
Here we determine the norm of the projected hole excitations: 
 \begin{eqnarray} 
\dfrac{N_{k+K,\sigma}^-}{N_G} &=& \dfrac{\langle \Psi_0|c_{\kvec+\Kvec,\sigma}^{\dagger}P Pc^{\phantom{\dagger}}_{\kvec+\Kvec,\sigma}|\Psi_0\rangle}{\langle \Psi|\Psi\rangle} \ .
\label{holnorm}
\end{eqnarray}
After the due computations in the GA, we obtain 
\begin{eqnarray}
\dfrac{N_{k+K,\sigma}^-}{N_G} = n_\sigma + \dfrac{1}{L}\sum_{l\gamma}\left[\dfrac{1-X_{\gamma}}{X_{\gamma}}\right] + \dfrac{1}{L}\sum_{lm\gamma\delta}\left[\left({\rm e}^{i\kvec \cdot (l-m)}{\rm e}^{i(\kvec + \Kvec) \cdot (\gamma-\delta)} \langle c_{l\gamma\sigma}^\dagger c_{m\delta\sigma}^{\phantom{\dagger}}\rangle_{0} - n^0_{l\gamma\sigma}\delta_{lm}\delta_{\gamma\delta}\right)Z_{\gamma\delta\sigma}\right]\label{NMZ} \ , 
\end{eqnarray}
with
\begin{eqnarray}
Z_{\gamma\delta\sigma} = \dfrac{1}{\sqrt{(1-n_{l\gamma\sigma})}}\left[\sqrt{(1-n_{l\gamma})}+\sqrt{\dfrac{n_{l\gamma,-\sigma}}{n_{l\gamma\sigma}}\cdot \dfrac{1-X_{\gamma}}{X_{\gamma}}}\right] \cdot \dfrac{1}{\sqrt{(1-n_{m\delta\sigma})}}\left[\sqrt{(1-n_{m\delta})}+ \sqrt{\dfrac{n_{m\delta,-\sigma}}{n_{m\delta\sigma}}\cdot \dfrac{1-X_{\delta}}{X_{\delta}}}\right] 
\label{Zgen}
\end{eqnarray}
where we remark that $l$ ($m$) belongs to the Bravais lattice and $\gamma$ ($\delta$) specifies the position in the unit cell and then the sublattice (see Sect.~II). 
Finally, we find  
\begin{eqnarray}
\dfrac{N_{k\sigma}^-}{N_G} &=&  n_\sigma + \dfrac{1-X}{X} + \left[Z_{AB\sigma}(u^2_{\kvec}n^{0}_{\kvec1\sigma} + v^2_{\kvec}n^{0}_{\kvec2\sigma})  - \dfrac{1}{2}Z_{AA\sigma}n^0_{A\sigma} - \dfrac{1}{2}Z_{AA,-\sigma}n^0_{A,-\sigma}\right]\ \nonumber \\
\dfrac{N_{k+K_{AF},\sigma}^-}{N_G} &=& n_\sigma + \dfrac{1-X}{X}  + \left[Z_{AB\sigma}(v^2_{\kvec}n^{0}_{\kvec 1\sigma} + u^2_{\kvec}n^{0}_{\kvec 2\sigma})  - \dfrac{1}{2}Z_{AA,-\sigma}n^0_{A,-\sigma} - \dfrac{1}{2}Z_{AA\sigma}n^0_{A\sigma}\right]
\label{NmZAFM} 
\end{eqnarray}
\begin{figure}
\includegraphics[width=8.65cm]{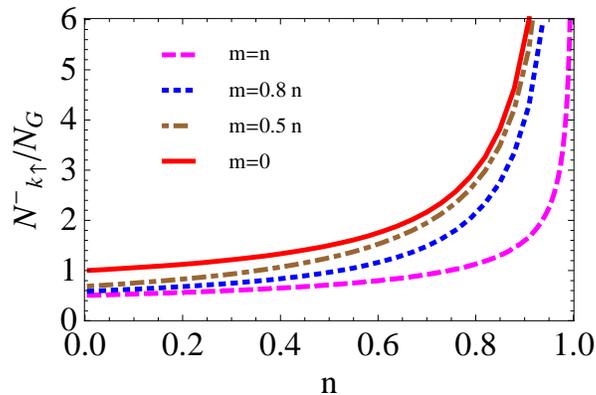}
\caption{Ratio between the hole excitation norm $N^-_{\kvec\uparrow}$ and the norm $N_G$ of the correlated state $\Psi$,  as a function of the filling $n$, for several magnetizations $m$.}
\label{Nhn}
\end{figure}
where 
the $Z$-factors are defined in Eq.~(\ref{Zgen}). 
\end{widetext}
For example, in Fig.~\ref{Nhn} we report the behaviour of the ratio between the hole excitation norm $N_{k\uparrow}^-$ and the norm $N_G$ of the correlated state $\Psi$, as a function of the electronic filling $n$ for some relative magnetizations $m/n$.
So we learn when projected hole excited states are possible and how the removal of one electron from the ground state is favoured or hindered by the change of the possible parameters $n$ and $m$. If compared to our previous findings with $N_{k\uparrow}^+ /N_G$, we can notice that the behaviour of $N_{k\uparrow}^- /N_G$ is only moderately affected by the choice of the parameters, except close to half-filling. In fact, upon increasing the magnetization, the probability for a hole excitation decreases moderately. At half-filling all the curves for different $m/n$ diverge, as we could expect thanks to the functional form of Eq.~(\ref{holnorm}) and the knowledge that the $X$-factor is 1 at half-filling for any relative magnetization. The trends revealed in Fig.~\ref{Npn} and \ref{Nhn} are opposite upon increasing $m/n$: the probability for projected particle-excited states is enhanced and the one for projected hole-excited states is reduced.

\section{Tunnelling probability}\label{sec:tunnel}

\subsection{Particle tunnelling}\label{subsec:partunnel}
Now we consider the tunnelling probability of a single particle into the projected state: 
\begin{equation}
M_{k+K,\sigma}^+ \ =\ 
\dfrac{\left| \langle \Psi_{k+K,\sigma}^+| 
 c_{\kvec+\Kvec,\sigma}^\dagger |\Psi\rangle  \right|^2}{N_{k+K,\sigma}^+ N_G} \ .
\label{M+}
\end{equation}
Since
\begin{eqnarray*}
&&\dfrac{\langle \Psi_{k+K,\sigma}^+| 
 c_{\kvec+\Kvec,\sigma}^\dagger |\Psi \rangle}{N_G} = \dfrac{\langle \Psi_0| c_{\kvec+\Kvec,\sigma}^{\phantom{\dagger}} P
 c_{\kvec+\Kvec,\sigma}^\dagger P |\Psi_0\rangle}{N_G} \\ &=& \dfrac{\langle \Psi_0| c_{\kvec+\Kvec,\sigma}^{\phantom{\dagger}} P
  P c_{\kvec+\Kvec,\sigma}^\dagger |\Psi_0\rangle}{N_G} 
 \equiv \dfrac{N_{k+K,\sigma}^+}{N_G} \ , 
\label{M+num}
\end{eqnarray*}
we find that
\begin{eqnarray} 
M_{k+K,\sigma}^+  = \dfrac{N_{k+K,\sigma}^+}{N_G} \ . 
\label{M++}
\end{eqnarray}
Eq.~(\ref{M++}) implies that the particle tunnelling probability has the same properties described for the particle excitation in Sect.~III A: it is enhanced for any fillings upon increasing the magnetization, but it is always forbidden at half-filling for any magnetizations.  

\subsection{Hole tunnelling}\label{subsec:holtunnel}

Here we evaluate the tunnelling probability of a single hole into the projected state: 
\begin{equation}
M_{k + K,\sigma}^- \ =\ 
\dfrac{\left| \langle \Psi_{k + K,\sigma}^-| 
 c_{\kvec + \Kvec,\sigma} |\Psi\rangle  \right|^2}{N_{k + K,\sigma}^- N_G} \ . 
\label{M-}
\end{equation}
After the analytical derivation reported in Appendix~1D, we find
\begin{widetext}
\begin{eqnarray}
M_{k+K,\sigma}^- &=& \dfrac{1}{L}\dfrac{\left|\sum_{l\gamma, m\delta}\left[(n_{l\gamma\sigma} - n^0_{l\gamma\sigma}Y_{\gamma\delta\sigma})\delta_{lm}\delta_{\gamma\delta}+ {\rm e}^{i (\kvec + \Kvec)\cdot (\gamma-\delta)}{\rm e}^{i \kvec \cdot (l-m)}\langle c_{l\gamma\sigma}^\dagger c_{m\delta\sigma}^{\phantom{\dagger}}\rangle_0 Y_{\gamma\delta\sigma}\right]\right|^2}{\sum_{l\gamma}\left[n_{l\gamma\sigma}+\dfrac{1-X_{\gamma}}{X_{\gamma}}\right] + \sum_{l\gamma,m\delta}\left[\left({\rm e}^{i\kvec \cdot (l-m)}{\rm e}^{i(\kvec + \Kvec) \cdot (\gamma-\delta)} \langle c_{l\gamma\sigma}^\dagger
c_{m\delta\sigma}^{\phantom{\dagger}}\rangle_{0} - n^0_{l\gamma\sigma}\delta_{lm}\delta_{\gamma\delta}\right)Z_{\gamma\delta\sigma}\right]} \ , 
\label{MmY}
\end{eqnarray}
with
\begin{eqnarray}
Y_{\gamma\delta\sigma} = \dfrac{\left[\sqrt{n_{l\gamma,-\sigma}}\sqrt{\dfrac{1-X_{\gamma}}{X_{\gamma}n_{l\gamma\sigma}}}+\sqrt{(1-n_{l\gamma})} \right]\sqrt{(1-n_{m\delta})}}{\sqrt{(1-n_{l\gamma\sigma})(1-n_{m\delta\sigma})}} \ . 
\label{Ylm}
\end{eqnarray}
For the AFM, Eq.~(\ref{MmY}) becomes
\begin{eqnarray}
M_{k\sigma}^- &=& \dfrac{\left|n_\sigma -\dfrac{1}{2}Y_{AA\sigma} n^0_{A\sigma} - \dfrac{1}{2}Y_{AA,-\sigma} n^0_{A,-\sigma} + Y_{AB\sigma}(u^2_{\kvec}n^{0}_{\kvec1\sigma} + v^2_{\kvec}n^{0}_{\kvec2\sigma})\right|^2}{n_\sigma + \dfrac{1-X}{X} + \left[Z_{AB\sigma}(u^2_{\kvec}n^{0}_{\kvec1\sigma} + v^2_{\kvec}n^{0}_{\kvec2\sigma}) - \dfrac{1}{2}Z_{AA\sigma}n^0_{A\sigma} - \dfrac{1}{2}Z_{AA,-\sigma}n^0_{A,-\sigma}\right]}  \nonumber \\ 
M_{k+K_{AF},\sigma}^- &=& \dfrac{\left|n_\sigma -\dfrac{1}{2}Y_{AA,-\sigma}n^0_{A,-\sigma} - \dfrac{1}{2}Y_{AA\sigma} n^0_{A\sigma} + Y_{AB\sigma}(v^2_{\kvec}n^{0}_{\kvec 1\sigma} + u^2_{\kvec}n^{0}_{\kvec 2\sigma})\right|^2}{n_\sigma + \dfrac{1-X}{X} + \left[Z_{AB\sigma}(v^2_{\kvec}n^{0}_{\kvec 1\sigma} + u^2_{\kvec}n^{0}_{\kvec 2\sigma}) - \dfrac{1}{2}Z_{AA,-\sigma}n^0_{A,-\sigma} - \dfrac{1}{2}Z_{AA\sigma}n^0_{A\sigma}\right]} \ , 
\label{MmYAFM}
\end{eqnarray}
where 
the $Z$-factors are defined in Eq.~(\ref{Zgen}) and the $Y$-factors in Eq.~(\ref{Ylm}). 
\end{widetext}
\begin{center}
\begin{figure}[bp]
\includegraphics[width=8.65cm]{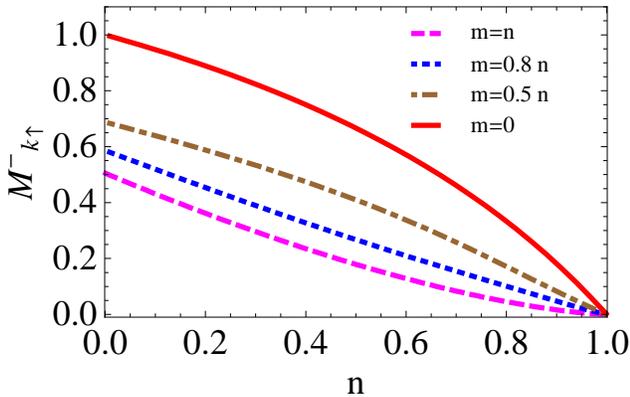}
\caption{Hole tunnelling probability $M^-_{\kvec\uparrow}$ as a function of the filling $n$, for several magnetizations $m$.}
\label{Mhn}
\end{figure}
\end{center}
For example, in Fig.~\ref{Mhn}  we report the behaviour of the hole tunnelling probability $M_{k\uparrow}^-$ as a function of the electronic filling $n$ for different relative magnetizations $m/n$.
This figure illustrates the behaviour of the tunnelling probability of one electron extracted from the ground state, upon changing the filling $n$ for some values $m/n$.  
In general we notice a decrease of the hole tunnelling probability, more pronounced for the N\'eel state. At half-filling both the hole tunnelling and the particle tunnelling are forbidden. This property of the half-filled AFM was expected for the introduction of a particle into the system and occurs also for his removal, displaying a particle-hole symmetry.  Upon increasing the magnetization, there is an enhancement of the tunnelling probabilities on the particle side and a reduction on the hole side with respect to the homogeneous case presented in Refs.~\onlinecite{Fuk05,Ede07}. 

The particle and hole matrix elements presented in Eqs.~(\ref{M++}) and (\ref{MmYAFM}) contribute to the conductance probed in the tunnelling experiments:\cite{Wie09} this illustrates the interest to work with the plane wave excitations $|\Psi_{k+K,\sigma}^\pm\rangle$. In fact, the excitations in Eqs.~(\ref{psi+}) and (\ref{psi-}) are responsible for the processes in tunnelling experiments, where the injected/extracted electrons are travelling waves.

However, one can also be interested to perform a similar study, considering the projected Bogolubov excitations, {\em i.e.} the quasiparticle excitation
\begin{equation}
|\Psi_{k\lambda\sigma}^+\rangle  \,=\, P \alpha_{\kvec\lambda\sigma}^\dagger|\Psi_0\rangle \ , 
\label{psil+}
\end{equation}
and the quasihole excitation
\begin{equation}
|\Psi_{k\lambda\sigma}^-\rangle  \,=\, P \alpha_{\kvec\lambda\sigma}|\Psi_0\rangle.
\label{psil-}
\end{equation}
The excitations in Eqs.~(\ref{psil+}) and (\ref{psil-}) are associated to the electronic eigenstates and occur in photoemission experiments. 

For the excitations $|\Psi_{k\lambda\sigma}^\pm\rangle$, one can evaluate the norms $N_{k\lambda\sigma}^\pm=\langle \Psi_{k\lambda\sigma}^\pm |\Psi_{k\lambda\sigma}^\pm\rangle$ and the tunnelling probabilities  
\begin{eqnarray*}
M_{k\lambda\sigma}^+  &=& 
\dfrac{\left| \langle \Psi_{k\lambda\sigma}^+| 
 \alpha_{\kvec\lambda\sigma}^\dagger |\Psi\rangle  \right|^2}{N_{k\lambda\sigma}^+ N_G}  \ , \nonumber \\ 
M_{k\lambda\sigma}^- &=& 
\dfrac{\left| \langle \Psi_{k\lambda\sigma}^-| 
 \alpha_{\kvec\lambda\sigma} |\Psi\rangle  \right|^2}{N_{k\lambda\sigma}^- N_G} \ . 
\label{Ml+-}
\end{eqnarray*}
The latter quantities can be presented using the results formerly obtained for the excitations $|\Psi_{k+K,\sigma}^\pm\rangle$:
\begin{eqnarray}
N_{k1\sigma}^\pm &=& u^2_\kvec N^\pm_{\kvec\sigma} + v^2_\kvec N^\pm_{\kvec+\Kvec_{AF},\sigma} \nonumber \\ 
N_{k2\sigma}^\pm &=& v^2_\kvec N^\pm_{\kvec\sigma} + u^2_\kvec N^\pm_{\kvec+\Kvec_{AF},\sigma} \ ; 
\end{eqnarray}
\begin{widetext}
\begin{eqnarray}
M_{k\lambda\sigma}^+ &=& N_{k\lambda\sigma}^+/N_G  \nonumber \\
M_{k1\sigma}^- &=& 
\dfrac{\left| \langle \Psi_{0}| 
 u^2_\kvec c^\dagger_{\kvec\sigma} P c^{\phantom{\dagger}}_{\kvec\sigma} P + v^2_\kvec  c^\dagger_{\kvec+\Kvec_{AF},\sigma} P c^{\phantom{\dagger}}_{\kvec+\Kvec_{AF},\sigma} P|\Psi_0\rangle \right|^2 }{N_{k1\sigma}^- N_G} \nonumber \\
M_{k2\sigma}^- &=& 
\dfrac{\left|  \langle \Psi_{0}| v^2_\kvec
 c^\dagger_{\kvec\sigma} P c^{\phantom{\dagger}}_{\kvec\sigma} P + u^2_\kvec c^\dagger_{\kvec+\Kvec_{AF},\sigma} P c^{\phantom{\dagger}}_{\kvec+\Kvec_{AF},\sigma} P|\Psi_0\rangle \right|^2}{N_{k2\sigma}^- N_G} \ . 
\label{NMl+-}
\end{eqnarray} 
\end{widetext}

\section{Monte Carlo Test}\label{sec:Montecarlo}

In order to generate the possible electronic configurations for the AFM states, we proceed with the following scheme in our VMC computations. 

We use the uncorrelated state $|\Psi_0\rangle$ obtained as the ground state of the mean-field Hamiltonian $H^0_{AF}$:
\begin{eqnarray}
H^0_{AF} = \sum_{i\alpha j\beta \sigma}t_{i\alpha j\beta}c^{\dagger}_{i\alpha\sigma}c^{\phantom{\dagger}}_{j\beta\sigma}-\Delta\sum_{i\alpha}(-1)^{i\alpha} S^z_{i\alpha} \ , 
\label{HAFM}
\end{eqnarray}
where $S^z_{i\alpha}$ is the $z$-component of the spin operator $\hat{S}$ on the lattice site $i\alpha$.

Then we consider the fully projected state $|\Psi\rangle$ and the partially projected state $|\Psi'_{l\gamma}\rangle$ [see Eqs. from (\ref{psi_l}) to (\ref{Ptot})] on a two-dimensional square lattice, with the nearest-neighbour hopping $t_{i\alpha j\beta}$ = $-t$. In this case the bare dispersion is $\epsilon^0_{\kvec\sigma} = -2t (\cos k_x + \cos k_y)$ and the vector $\Kvec_{AF} = (\pi,\pi)$.

In our VMC computations, we evaluated the double occupancy $d_{\gamma}$ of the unprojected reservoir site for a comparison with the corresponding GA results: this will provide a test for the accuracy of the GA for projected AFM wave functions. 
In our Monte Carlo sampling of the electronic configurations, two kinds of updates are allowed to generate a new configuration from the previous one: either one-electron hopping or spin-flip between two nearest-neighbour sites. 
The contribution of the spin-flip processes is more relevant close to half-filling and necessary at half-filling to update the configurations when double occupancies are not allowed, as in fully projected states $\Psi$. Both the types of update processes do not change the condition $N_\uparrow = N_\downarrow$ for the whole lattice but they allow to change the staggered magnetization of the microscopical configuration.  
Furthermore, since we need to evaluate an intensive local quantity ($d_\gamma$), we update the reservoir site oftener than the projected ones, in order to accumulate enough statistics in a shorter time.
The standard update procedure uses the equiprobability to select two sites and then the configuration update interests the reservoir site on average once every $L$ steps. If $L \approx 10^3$ the thermalization towards the equilibrium value will require a very long time for a local quantity. Instead, if we select the reservoir site $10$-$20$ times oftener, 
this will shorten the time to achieve equilibrium value but nevertheless the statistics accumulated for the other sites is enough to have an undistorted description also of the surrounding environment. In this way we obtain an unbiased value for the double occupancy of the unprojected site, preserving the local detailed balance condition. 

\begin{center}
\begin{figure}[bp]
\includegraphics[width=8.65cm]{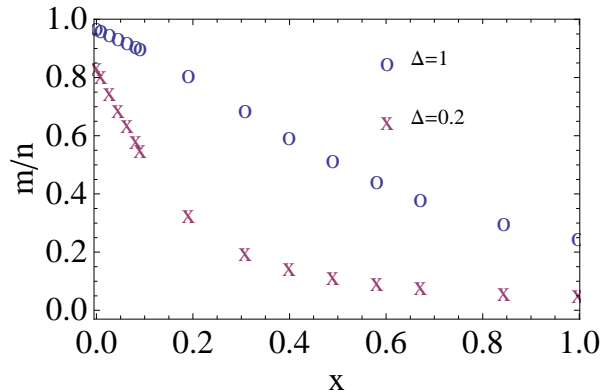}
\caption{Ratio between the magnetization $m$ and the filling $n$ as a function of the hole doping $x$ for two values of the staggering field $\Delta$ on a 882-site lattice (VMC results).}
\label{mnDelta}
\end{figure}
\end{center}

Our VMC computations were performed typically on lattices ranging from 882 to 2500 sites: the size-dependent effects are quite negligible for our investigations with AFM wave functions. In addition, we collected enough statistics in the number of electronic configurations generated for the sampling, so that the uncertainties of the data points here reported are within $1\%$, and then within the symbol size in the plots.  

In Fig.~\ref{mnDelta} we show the behaviour of the relative magnetization $m/n$, upon increasing the hole doping $x=1-n$: this trend changes for different $\Delta$, whose values are given in unit of the hopping parameter $t$. In our VMC approach, we obtain a state with a finite magnetization upon applying a finite external field $\Delta$. Thanks to this kind of study we find the whole correspondence between our control parameter $\Delta$ and the order parameter $m$ of the AFM states. In fact, we know that for $\Delta=0$ we recover the paramagnetic Fermi  sea ($m=0$), and for $\Delta=\infty$ we obtain the N\'eel state with $m/n=1$, while in the plot we present the response of the system for two intermediate values of $\Delta$. 
For small $\Delta$, $m/n$ drops quite immediately upon increasing the doping and then is few percents of the half-filled value for intermediate and large dopings. This trend takes places because a weak field $\Delta$ is effective to induce a strong magnetization only if the electronic density is high, otherwise the system does not tend to place electrons with opposite spin on either sublattices.  
For large $\Delta$, the decrease of the $m/n$ upon increasing the doping is quasi-linear for a wide range of dopings: in a strong field it is convenient to place some electrons according to the AFM order also when only few electrons are on the lattice.  
A final interesting issue emerges in this investigation: for any nonzero $\Delta$ in the low-density limit the relative magnetization does not vanish and tends to a finite value. This implies that even diluted systems, and possibly the lattice gases, show quite sizable magnetization for finite not necessarily huge $\Delta$: actually, the electronic system has a memory of the underlying bipartite lattice even if the vacancies exceed highly the occupied sites.  

\begin{center}
\begin{figure}[bp]
\includegraphics[width=8.65cm]{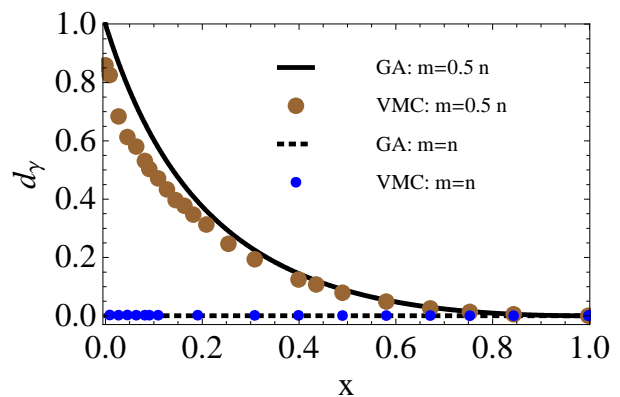}
\caption{Double occupancy $d_\gamma$ of the unprojected reservoir site as a function of the hole doping $x$ on a 882-site lattice: GA and VMC results are compared for intermediate magnetization $m$ and for the N\'eel state ($n$ is the filling).} 
\label{dgAFM}
\end{figure}
\end{center} 
Then we consider the double occupancy $d_\gamma$ of the reservoir site and we compare the results obtained in the GA, using Eqs.~(\ref{xguess}) and (\ref{psi_l_double}), and the VMC results. For $m=0$, we recover the results  investigated in Ref.~\onlinecite{Fuk05} for the projected Fermi sea, with an excellent agreement between the GA and the VMC for any dopings. 

In Fig.~\ref{dgAFM} we plot the double occupancy $d_\gamma$, as a function of the doping, for two magnetizations: $m=n$ (the N\'eel state) and $m=n/2$. 
For $m=n$, the probability to have a double occupancy is zero: in a very strong staggering field there is no chance to have two electrons on the same site, even on the reservoir one, and this is a behaviour shown both by uncorrelated magnetic systems and correlated ones. For intermediate staggering fields $\Delta$ and $m/n$, we find that the GA results are in an excellent accord with the numerical ones for intermediate and large dopings. Instead, the agreement is less satisfactory for small dopings and particularly poor very close to half-filling. In general, this is quite reasonable due to the nature of the GA, whose performances are not very good for dense systems. In fact, in the GA all the configurations with the same number of $\uparrow$- and $\downarrow$-electrons give the same contribution to the Hilbert space, thus completely neglecting their spatial locations on the lattice. Then the GA does its best for homogeneous states and in general for not too dense systems: not so close to half-filling, the mutual interactions among the electrons on the lattice are quite sporadic and their specifical positions are not so crucial.   

The comparison of our results in GA with the corresponding VMC ones confirms the accuracy of the approximation. This is in general satisfactory for intermediate and large dopings and at least reliable for the qualitative trends for low dopings, where in any case the discrepancy is within  $20\%$ with respect to the numerical values. The agreement is worse for intermediate magnetizations.  

A similar study was performed in Ref.~\onlinecite{Fuk05} for the projected Fermi sea and for projected BCS-superconducting states. Since some open questions were left for the projected superconducting states, we decided to perform the computations in Appendix~2, in order to clarify and possibly fix these issues. 

\section{Discussion}\label{sec:discuss}

The determination of the tunnelling matrix elements $M^{\pm}_{k}$ presented in Sect.~IV is relevant to study the results of one-particle spectroscopies, as STM. 

These matrix elements are very difficult to be evaluated in an accurate realistic way, both analytically and numerically. 
The GA allows us to have insight into the structure of $M^{\pm}_{k}$, and the accuracy of the approximation is tested in Sect.~\ref{sec:Montecarlo}, where we compare analytical results in the GA and exact numerical VMC results (see for example Fig.~\ref{dgAFM}). The discrepancy between the approximated and the exact results is very tiny for moderate and large dopings. The worst agreement is found close to half-filling for intermediate magnetizations, where the deviation is up to 20$\%$. On the whole, the reliability of the approximation is satisfactory, and even very good for most investigated cases.         

Previous papers evaluated tunnelling probabilities for homogeneous states, mainly to study the spectral properties of high-temperature superconductors.\cite{Par01, Yun06, Cap06, Nav06, Bie07,Ede06} Thus we decided to explore the tunnelling for the AFM states, not investigated yet.  

Tunnelling experiments allow for a complete study of one-particle processes: both the particle and the hole side of the spectrum can be obtained in the same experimental setup, just inverting the potential difference between the metallic tip and the sample. Unfortunately, this kind of method can 
not be used for insulators because the potential barrier would be too large to have an acceptably high signal-to-noise ratio. The absence of STM spectra for insulators hinders us to have experiments for a direct comparison with our results. In principle, we need experimental results for correlated collinear AFM materials, not necessarily for insulators. But, in practice, all the strongly correlated materials which exhibit this specifical magnetic order are insulators, as the lowly doped cuprates.\cite{Nag97, Lee06}  
Other materials with moderate electronic correlations are AFM metals, as the iron-pnictides, but in this case the magnetic order is a stripe-like SDW.\cite{Ish09} Our findings of Sect.~III and IV could possibly interest other novel correlated materials. 

Single photoemission experiments can not  describe both the hole and the particle side of the spectrum: in fact, hole excitations occur in direct photoemission, when the material is hit by photons and emits electrons; instead particle excitations take place in inverse photoemission, when  the material is irradiated by an electronic beam and emits photons.\cite{Dama03}    

An important feature in tunnelling spectra for correlated materials is the particle-hole asymmetry.\cite{Hir99, Pan00, Ran00, Han04, Ran04, Par05, And04, And06, Yan07, Yan10} 
In uncorrelated materials the spectral weight for the excitations is symmetrical in its particle and hole side: this implies that the probability to add or remove one electron is the same. On the contrary, in correlated materials a particle-hole asymmetry is detected and the hole-side of the spectrum dominates. This is due to the strongly correlated environment which penalises the introduction of an extra electron into the system but not its removal. 
Actually,  the tunnelling spectra show only a small asymmetry for low-energy excitation, being the most asymmetric features for high-energy excitations.
Theoretically, the asymmetry was investigated so far especially 
for correlated superconductors. The mostly accepted idea is that low-energy coherent excitations are responsible for a particle-hole symmetry in the spectrum also in correlated systems, and then the whole feature of the asymmetry would be due to the high-energy incoherent excitations. This scenario is illustrated by Randeria {\em et al.}\cite{Ran04} using projected BCS wave functions and a sum rule analysis for models with coulombic correlations; similar conclusions are also the outcome of other dedicated studies.\cite{Ran00, Par05, Yan07, Yan10} The complementary idea is that a mean-field treatment of the electronic excitations can not capture any incoherent contribution to the total spectrum and then it would miss the asymmetry at all, as found in the explicit computations with the projected Gutzwiller excited states in homogeneous systems.\cite{Fuk05} 
A different position is expressed by Anderson and Ong\cite{And04, And06} and this is relevant for our present study. They deny that the Gutzwiller scheme for the projected quasiparticle excitations is intrinsically a mean-field approach limited to the coherent part of the spectrum and not able to find any asymmetry. Then in their approach they formulate projected excitations and obtain the particle-hole asymmetry in the tunnelling. This achievement is somehow in contradiction with other systematical investigations as in Ref.~\onlinecite{Fuk05}.

The previous studies are mainly motivated by the interest in the superconducting phase of materials like the cuprates. Instead, we decided to investigate the results of similar computations for projected AFM states. 

We point out that our framework is valid for low-energy particle (hole) excitations, the region where in the superconductors the tunnelling spectra exhibit only a small asymmetry.\cite{Pan00}

In Sect.~III and IV, we determined the general properties for the tunnelling elements, computed using the AFM projected wave functions defined by Eqs.~(\ref{psi}), (\ref{Ptot}), (\ref{AFPsi0}) and (\ref{uvk}) on a bipartite lattice, without further specifications of the model to be used.

The main outcome of our work is that the increase of the magnetization has opposite effects on the behaviour of the particle and hole excitations norms (as shown in Sect. III) and also of particle and hole tunnelling matrix elements (as shown in Sect. IV). In fact, upon increasing the magnetization from the paramagnetic state to the N\'eel state, we found that the particle tunnelling is enhanced and, on the contrary, the hole tunnelling is reduced though never suppressed. These changes in the tunneling are monotonous upon varying the magnetization. We emphasize that these findings are general for all the range of possible fillings. Therefore, this issue is expected to be robust, so to allow us for the following prediction: if one compares the tunnelling spectra of correlated materials with collinear AFM order, one should find that the ratio of the particle-side vs hole-side signal has to increase with the magnetization of the system. Actually, other effects contribute to give the conductance and they can possibly hide or alter this trend shown by the tunnelling matrix elements\cite{Fis07}; however, for materials with quite different magnetizations, the effect should be clearly observable in the low-energy sector of the spectra. 

Finally, the general form of our results in Eqs.~(\ref{N++}), (\ref{M++}), (\ref{MmYAFM}) and (\ref{NMl+-}) is not explicitly particle-hole symmetrical.  
However, we identified relevant particle-hole symmetrical cases, as the homogeneous limit for any fillings or the half-filled case for any magnetizations.
 For practical purposes, correlated materials exhibit AFM order  close to half-filling and then for this range of fillings our results can just allow for a tiny deviation from the particle-hole symmetry.

\section{Conclusions}\label{sec:conc}

Our investigation provides the tunnelling matrix elements for projected AFM wave functions in the GA. Such matrix elements contribute to the signal intensity  in the one-electron spectroscopies for collinear AFM-ordered materials. To our knowledge, ours is the first theoretical work dedicated to study single-particle (hole) tunnelling for projected AFM states. 

We also tested the accuracy of our analytical results in GA through a comparison with VMC numerical results, findind a generally good agreement for any fillings and magnetizations.  

We explored the behaviour of the particle (hole) excitation norms and tunnelling probabilities for the fillings up to half-filling and for the whole range of possible magnetizations. The main outcome is that an increasing magnetization enhances the particle tunnelling and reduces the hole tunnelling. Then, we also identified a particle-hole symmetrical behaviour of the tunnelling for some relevant cases.  

For future investigations, we are developing a natural extension of our present study and of the one in Ref.~\onlinecite{Fuk05}, evaluating matrix elements for projected two-particle (hole) excitations. This further analysis will be helpful to better understand the classes of experiments which probe a response function of the system, as for example charge and spin susceptibility in optical and neutron scattering, optical conductivity and resistivity.   

\acknowledgments

The authors acknowledge the support of the German Science Foundation (DFG).

\

\section*{Appendix 1}\label{sec:appendixA}

\subsection{Gutzwiller factors for the AFM}\label{subsec:GAfact}

The general expression for the hopping factor is\cite{Ede07} 
\begin{eqnarray}
g_{t\alpha\beta\uparrow} &=& \dfrac{\langle c^\dag_{i\alpha\uparrow}c_{j\beta\uparrow}\rangle}{\langle c^\dag_{i\alpha\uparrow}c_{j\beta\uparrow}\rangle_0} \nonumber \\ &=& 
\dfrac{[n_{i\alpha\uparrow}(1-n_{j\beta})n_{j\beta\uparrow}(1-n_{i\alpha})]^{1/2}}{[n^0_{i\alpha\uparrow}(1-n^0_{j\beta\uparrow})n^0_{j\beta\uparrow}(1-n^0_{i\alpha\uparrow})]^{1/2}}.
\label{gtij}
\end{eqnarray}
For a SDW we know that  
$n_{i\alpha} =  n_{j\beta} = n^0_{i\alpha} = n^0_{j\beta} = n$ and we can also relate the projected spin density $n_{i\alpha\sigma}$ and the unprojected one, $n^0_{i\alpha\sigma}$:  
$n_{i\alpha\sigma} = C n^0_{i\alpha\sigma}(1-n^0_{i\alpha-\sigma})$, where $C$ is a constant which we obtain, noting that 

$n^0 = n^0_{i\alpha\sigma} + n^0_{i\alpha-\sigma} = n_{i\alpha\sigma} + n_{i\alpha-\sigma} = n$

and summing the following two equations: 
\begin{eqnarray*}
n_{i\alpha\sigma} &=& C n^0_{i\alpha\sigma}(1-n^0_{i\alpha-\sigma}) \\ 
n_{i\alpha-\sigma} &=& C n^0_{i\alpha-\sigma}(1-n^0_{i\alpha\sigma}) \\ 
C &=& \dfrac{1}{1-\dfrac{2n^0_{i\alpha\sigma} n^0_{-i\alpha\sigma}}{n}}.
\end{eqnarray*}
Then we have this relation for a SDW: 
\begin{eqnarray}
n_{i\alpha\sigma} =  \dfrac{1}{1-\dfrac{2n^0_{i\alpha\sigma} n^0_{i\alpha-\sigma}}{n}} n^0_{i\alpha\sigma}(1-n^0_{i\alpha-\sigma}).
\label{nsSDW}
\end{eqnarray}
We can replace the latter relation for $n_{i\alpha\sigma}$ in Eq.~(\ref{gtij}) and obtain the factor $g_t$ for a generic SDW, given in Eq.~(\ref{gSDW}).  

To formulate the Gutzwiller factors for an AFM, we can adopt this kind of shorthand notation: $n_{A\sigma}= n_{B,-\sigma} = n_{\sigma}$.
\begin{widetext}
In an AFM also the magnetization $m$ undergoes a renormalization with respect to the unprojected $m^0$.  Using Eq.~(\ref{nsSDW}), we find  
\begin{eqnarray*}
m &=& n_{i\uparrow} - n_{i\downarrow}  
= \dfrac{1}{1-\dfrac{2n^0_{i\sigma} n^0_{i-\sigma}}{n}}(n^0_{i\uparrow} - n^0_{i\downarrow}) = \dfrac{m^0}{1- \dfrac{n^2-(m^0)^2}{2n}} \ .
\label{mm0}
\end{eqnarray*}
If the sites $i\alpha$ and $j\beta$ are on different sublattices, one obtains the factor $g_{tAB} \equiv g_{tAB\sigma}$: 
\begin{eqnarray*} 
g_{tAB} &=& = \dfrac{1-n}{1-2 n^0_\uparrow n^0_\downarrow/n} = \dfrac{1-n}{1-\dfrac{n^2-(m^0)^2}{2n}} \ .
\end{eqnarray*}
If they are on the same sublattice, one obtains the factors $g_{tAA\sigma} \equiv g_{tA\sigma}$ and $g_{tBB\sigma} \equiv g_{tB\sigma}$: 
\begin{eqnarray*}
g_{tA\uparrow} &=& \dfrac{1-n}{1-2 n^0_\uparrow n^0_\downarrow/n}\dfrac{1-n^0_\downarrow}{1-n^0_\uparrow} = \dfrac{1-n}{1- \dfrac{n^2-(m^0)^2}{2n}} \dfrac{2-n+m^0}{2-n-m^0}  \ ,    \\ 
g_{tB\uparrow} &=& \dfrac{1-n}{1-2 n^0_\uparrow n^0_\downarrow/n}\dfrac{1-n^0_\uparrow}{1-n^0_\downarrow} = \dfrac{1-n}{1-\dfrac{n^2-(m^0)^2}{2n}}\dfrac{2-n-m^0}{2-n+m^0} \ .
\label{gtB}
\end{eqnarray*}
Note that $g_{tA\sigma} = g_{tB\,-\sigma}$.
\subsection{Particle excitation}\label{subsec:partexcit-append}
\begin{eqnarray} 
N_{k+K,\sigma}^+ \ &=&  
\langle \Psi_0|c_{\kvec+\Kvec,\sigma}^{\phantom{\dagger}}P Pc_{\kvec+\Kvec,\sigma}^\dagger|\Psi_0\rangle \nonumber \\  
\ &=& \dfrac{1}{L}
\sum_{l\gamma,m\delta}{\rm e}^{i \kvec \cdot(l-m)}{\rm e}^{i (\kvec+\Kvec) \cdot(\gamma-\delta)}
\langle \Psi_0|P_{l\gamma}' (1-n_{l\gamma,-\sigma})
c_{l\gamma\sigma}^{\phantom{\dagger}} c_{m\delta\sigma}^\dagger
(1-n_{m\delta,-\sigma}) P_{m\delta}'|\Psi_0\rangle 
\qquad \nonumber \\ \qquad \nonumber
 &=& \dfrac{1}{L} \left[\sum_{l\gamma} 
\langle \Psi_0|P_{l\gamma}'(1-n_{l\gamma\sigma})(1-n_{l\gamma,-\sigma})P_{l\gamma}'|\Psi_0\rangle + 
\sum_{l\gamma\ne m\delta} {\rm e}^{i \kvec \cdot(l-m)}{\rm e}^{i (\kvec+\Kvec) \cdot(\gamma-\delta)}\langle \Psi_0|P c_{l\gamma\sigma}^{\phantom{\dagger}} c_{m\delta\sigma}^\dagger
P|\Psi_0\rangle \right]\nonumber
\qquad \\ \qquad \label{partnorm-app}
\ &=& \dfrac{N_G}{L}\left[\sum_{l\gamma} 
\dfrac{\langle \Psi|(1-n_{l\gamma}) |\Psi\rangle}{\langle \Psi|\Psi\rangle }
\, + \,
\sum_{l\gamma\ne m\delta} {\rm e}^{i \kvec \cdot(l-m)}{\rm e}^{i (\kvec + \Kvec) \cdot(\gamma-\delta)} 
\dfrac{\langle \Psi|1-c_{m\delta\sigma}^\dagger c_{l\gamma\sigma}^{\phantom{\dagger}} |\Psi\rangle}{\langle \Psi|\Psi\rangle}\right] \ ,
\end{eqnarray}
where $N_G=\langle \Psi | \Psi \rangle$. For the diagonal contribution in the last step we have used the relation 

$\langle\Psi_0|P(1-n_{l\gamma})P|\Psi_0\rangle=\langle\Psi_0|P'_{l\gamma}(1-n_{l\gamma\uparrow})(1-n_{l\gamma\downarrow})P'_{l\gamma}|\Psi_0\rangle$.  
Then we invoke the GA for the off-diagonal term in Eq.~(\ref{partnorm-app}): 
\begin{eqnarray*}
\dfrac{N_{k+K,\sigma}^+}{N_G} &=& (1-n) - \dfrac{1}{L}\left[\langle\Psi_0|\sum_{l\gamma, m\delta}g_{t\gamma\delta\sigma} {\rm e}^{i \kvec \cdot(l-m)}{\rm e}^{i (\kvec +\Kvec) \cdot(\gamma-\delta)} c_{m\delta\sigma}^{\dagger} c_{l\gamma\sigma}^{\phantom{\dagger}} \right . \\ &-& \left .\sum_{(l,\gamma)=(m,\delta)} g_{t\gamma\delta\sigma}{\rm e}^{i \kvec \cdot(l-m)}{\rm e}^{i (\kvec +\Kvec) \cdot(\gamma-\delta)}c_{m\delta\sigma}^\dagger c_{l\gamma\sigma}^{\phantom{\dagger}}|\Psi_0\rangle\right] \\  
&=& (1-n) - \dfrac{1}{L}\left[\langle\Psi_0| - \sum_{l\gamma} g_{t\gamma\gamma\sigma} n^0_{l\gamma\sigma} \right . \\ &+& \left. \sum_{l\gamma, m\delta}g_{t\gamma\delta\sigma} {\rm e}^{i \kvec \cdot(l-m)}{\rm e}^{i (\kvec +\Kvec) \cdot(\gamma-\delta)} c_{m\delta\sigma}^{\dagger} c_{l\gamma\sigma}^{\phantom{\dagger}}|\Psi_0\rangle\right] \ . \nonumber
\label{norm_psi+}
\end{eqnarray*}
For the AFM the latter can be formulated in an explicit form thanks to Eq.~(\ref{Bog}): 
\begin{eqnarray*}
\dfrac{N_{k+K,\sigma}^+}{N_G} &=& (1-n) + \dfrac{1}{2}\left(g_{tA\sigma}n^0_{A\sigma}+g_{tA,-\sigma}n^0_{A,-\sigma}\right) - g_{tAB}\langle c^{\dagger}_{\kvec+\Kvec,\sigma}c^{\phantom{\dagger}}_{\kvec+\Kvec,\sigma}\rangle_{\Psi_0}.  
\label{norm_psiAFM+app}
\end{eqnarray*}
\subsection{Hole excitation }\label{subsec:holexcit-append}
\begin{eqnarray*} 
\dfrac{N_{k+K,\sigma}^-}{N_G} &=& \dfrac{\langle \Psi_0|c_{\kvec+\Kvec,\sigma}^{\dagger}P Pc^{\phantom{\dagger}}_{\kvec+\Kvec,\sigma}|\Psi_0\rangle}{\langle \Psi|\Psi\rangle} = 
\dfrac{1}{N_G L} \sum_{l\gamma,m\delta} {\rm e}^{i\kvec \cdot (l-m)}e^{i(\kvec+\Kvec)\cdot(\gamma-\delta)}\langle \Psi_0|P'_{l\gamma} c_{l\gamma\sigma}^\dagger
c_{m\delta\sigma}^{\phantom{\dagger}} P'_{m\delta}|\Psi_0\rangle \\ 
 &=&\dfrac{1}{L}\sum_{l\gamma}\dfrac{1}{X_{\gamma}}\Big[X_{\gamma}n_{l\gamma\sigma} + (1-X_{\gamma})\Big] + \dfrac{1}{N_G L} \sum_{l\gamma\ne m\delta} {\rm e}^{i\kvec \cdot (l-m)}{\rm e}^{i(\kvec + \Kvec) \cdot (\gamma-\delta)} \langle \Psi_0|P_{lm\gamma\delta}' c_{l\gamma\sigma}^\dagger c_{m\delta\sigma}^{\phantom{\dagger}} P_{lm\gamma\delta}'|\Psi_0\rangle \\
&=& n_\sigma + \dfrac{1}{L}\sum_{l\gamma}\left[\dfrac{1-X_{\gamma}}{X_{\gamma}}\right]  \\ &+& \dfrac{1}{N_G L} \left[\sum_{l\gamma, m\delta} {\rm e}^{i\kvec \cdot (l-m)}{\rm e}^{i(\kvec + \Kvec) \cdot (\gamma-\delta)}\langle \Psi'_{lm\gamma\delta}| c_{l\gamma\sigma}^\dagger c_{m\delta\sigma}^{\phantom{\dagger}}|\Psi'_{lm\gamma\delta}\rangle - \sum_{l\gamma} \langle \Psi'_{ll\gamma\gamma}|c_{l\gamma\sigma}^\dagger c_{l\gamma\sigma}^{\phantom{\dagger}}|\Psi'_{ll\gamma\gamma}\rangle \right] \ , 
\end{eqnarray*}
where $P'_{lm\gamma\delta} = \prod_{i\alpha \ne l\gamma,m\delta}(1-n_{i\alpha,\uparrow}n_{i\alpha,\downarrow})$.  
The last term in the above equation corresponds to a hopping
process between two reservoir sites. The generalized GA assumes that the matrix elements are proportional to the square roots of the corresponding densities, as given in Eq.~(\ref{psi_l_empty}). 
Invoking the GA, we obtain:
\begin{eqnarray*}
\dfrac{N_{k+K,\sigma}^-}{N_G} &=& n_\sigma + \dfrac{1}{L}\sum_{l\gamma}\left[\dfrac{1-X_{\gamma}}{X_{\gamma}}\right] + \dfrac{1}{L} \Big[\sum_{l\gamma, m\delta}\dfrac{1}{X_{\gamma}X_{\delta}} {\rm e}^{i\kvec \cdot (l-m)}{\rm e}^{i(\kvec + \Kvec) \cdot (\gamma-\delta)} \langle c_{l\gamma\sigma}^\dagger c_{m\delta\sigma}^{\phantom{\dagger}}\rangle_{0} \cdot  \\  &\cdot&  \dfrac{1}{\sqrt{n_{l\gamma\sigma}(1-n_{l\gamma\sigma})}}\left[\sqrt{X_{\gamma} (1-n_{l\gamma})}\sqrt{X_{\gamma} n_{l\gamma\sigma}}+\sqrt{X_{\gamma} n_{l\gamma,-\sigma}} \sqrt{1-X_{\gamma}}\right]   \\ &\cdot& \dfrac{1}{\sqrt{n_{m\delta\sigma}(1-n_{m\delta\sigma})}}\left[\sqrt{X_{\delta} (1-n_{m\delta})} \sqrt{X_{\delta} n_{m\delta\sigma}}+\sqrt{X_{\delta} n_{m\delta,-\sigma}} \sqrt{1-X_{\delta}}\right]  \\ &-&  \sum_{l\gamma} \dfrac{1}{X^2_{\gamma}}\dfrac{n^0_{l\gamma\sigma}}{n_{l\gamma\sigma}(1-n_{l\gamma\sigma})} \left[\,\sqrt{X_{\gamma} (1-n_{l\gamma})} \sqrt{X_{\gamma} n_{l\gamma\sigma}}+\sqrt{X_{\gamma}n_{l\gamma,-\sigma}} \sqrt{1-X_{\gamma}}\Big]^2 \right] 
\end{eqnarray*}
\begin{eqnarray}
\dfrac{N_{k+K,\sigma}^-}{N_G} = n_\sigma + \dfrac{1}{L}\sum_{l\gamma}\left[\dfrac{1-X_{\gamma}}{X_{\gamma}}\right] + \dfrac{1}{L}\sum_{lm\gamma\delta}\left[\left({\rm e}^{i\kvec \cdot (l-m)}{\rm e}^{i(\kvec + \Kvec) \cdot (\gamma-\delta)} \langle c_{l\gamma\sigma}^\dagger c_{m\delta\sigma}^{\phantom{\dagger}}\rangle_{0} - n^0_{l\gamma\sigma}\delta_{lm}\delta_{\gamma\delta}\right)Z_{\gamma\delta\sigma}\right]\label{NmZ} \ , 
\end{eqnarray}
with
\begin{eqnarray*}
Z_{\gamma\delta\sigma} = \dfrac{1}{\sqrt{(1-n_{l\gamma\sigma})}}\left[\sqrt{(1-n_{l\gamma})}+\sqrt{\dfrac{n_{l\gamma,-\sigma}}{n_{l\gamma\sigma}}\cdot \dfrac{1-X_{\gamma}}{X_{\gamma}}}\right] \cdot \dfrac{1}{\sqrt{(1-n_{m\delta\sigma})}}\left[\sqrt{(1-n_{m\delta})}+ \sqrt{\dfrac{n_{m\delta,-\sigma}}{n_{m\delta\sigma}}\cdot \dfrac{1-X_{\delta}}{X_{\delta}}}\right] \ . 
\label{Zgenapp}
\end{eqnarray*}
\subsection{Hole tunnelling}\label{subsec:holtunnel-append}

Here we evaluate the tunnelling probability of a single hole into the projected state: 
\begin{equation}
M_{k + K,\sigma}^- \ =\ 
\dfrac{\left| \langle \Psi_{k + K,\sigma}^-| 
 c_{\kvec + \Kvec,\sigma} |\Psi\rangle  \right|^2}{N_{k + K,\sigma}^- N_G} \ .
\label{M-app}
\end{equation}
We use Eq.~(\ref{psi_l_single}) and find:
\begin{eqnarray} 
&& \dfrac{\langle \Psi_0| c_{\kvec + \Kvec,\sigma}^\dagger P c_{\kvec + \Kvec,\sigma}^{\phantom{\dagger}} P |\Psi_0\rangle}{N_G} =  \dfrac{1}{N_G L}\sum_{l\gamma,m\delta} {\rm e}^{i (\kvec + \Kvec)\cdot (\gamma-\delta)}{\rm e}^{i \kvec \cdot (l-m)} \langle \Psi_0|P_{l\gamma}' c_{l\gamma\sigma}^\dagger c_{m\delta\sigma}^{\phantom{\dagger}} P|\Psi_0\rangle  \nonumber \\
&=& \dfrac{1}{N_G L}\left[\sum_{l\gamma} 
\langle \Psi'_{l\gamma}|c_{l\gamma\sigma}^\dagger c_{l\gamma\sigma}^{\phantom{\dagger}}|\Psi'_{l\gamma}\rangle + \sum_{l\gamma \neq m\delta}{\rm e}^{i (\kvec + \Kvec)\cdot (\gamma-\delta)}{\rm e}^{i \kvec \cdot (l-m)} \langle \Psi'_{l\gamma}|c_{l\gamma\sigma}^\dagger c_{m\delta\sigma}^{\phantom{\dagger}}|\Psi'_{l\gamma}\rangle \right] \nonumber \\ &=&
\dfrac{1}{N_G L}\left[\sum_{l\gamma}\dfrac{X_{\gamma}n_{l\gamma\sigma}}{X_{\gamma}}
\,+ \, \sum_{l\gamma, m\delta}{\rm e}^{i (\kvec + \Kvec)\cdot (\gamma-\delta)}{\rm e}^{i \kvec \cdot (l-m)} \langle \Psi'_{l\gamma}|c_{l\gamma\sigma}^\dagger c_{m\delta\sigma}^{\phantom{\dagger}}|\Psi'_{l\gamma}\rangle- \sum_{(l,\gamma)=(m,\delta)}\langle \Psi'_{l\gamma}|c_{l\gamma\sigma}^\dagger c_{m\delta\sigma}^{\phantom{\dagger}}|\Psi'_{l\gamma}\rangle\right] \nonumber \\
 &=& n_{\sigma} + \dfrac{1}{L}\left [\sum_{l\gamma, m\delta}\left[{\rm e}^{i (\kvec + \Kvec)\cdot (\gamma-\delta)}{\rm e}^{i \kvec \cdot (l-m)}\langle c_{l\gamma\sigma}^\dagger c_{m\delta\sigma}^{\phantom{\dagger}}\rangle_0 -n^0_{l\gamma\sigma}\delta_{lm}\delta_{\gamma\delta}\right] \cdot  \right . \\ &\cdot& \left .
\dfrac{\left[\sqrt{X_{\gamma}n_{l\gamma,-\sigma}}\sqrt{1-X_{\gamma}}
+\sqrt{X_{\gamma}(1-n_{l\gamma})}\sqrt{X_{\gamma}n_{l\gamma\sigma}} \right]\left[
\sqrt{1-n_{m\delta}}\sqrt{n_{m\delta\sigma}}\right]}{X_{\gamma} \sqrt{(1-n_{l\gamma\sigma})n_{l\gamma\sigma}}\sqrt{(1-n_{m\delta\sigma})n_{m\delta\sigma}}}\right]
\nonumber \\  \qquad \nonumber  
\  &=& n_{\sigma} + \dfrac{1}{L}\sum_{l\gamma, m\delta}\left[{\rm e}^{i (\kvec + \Kvec)\cdot (\gamma-\delta)}{\rm e}^{i \kvec \cdot (l-m)}\langle c_{l\gamma\sigma}^\dagger c_{m\delta\sigma}^{\phantom{\dagger}}\rangle_0 -n^0_{l\gamma\sigma}\delta_{lm}\delta_{\gamma\delta}\right] Y_{\gamma\delta\sigma}\nonumber 
\end{eqnarray}
\begin{eqnarray}
M_{k+K,\sigma}^- &=& \dfrac{1}{L}\dfrac{\left|\sum_{l\gamma, m\delta}\left[(n_{l\gamma\sigma} - n^0_{l\gamma\sigma}Y_{\gamma\delta\sigma})\delta_{lm}\delta_{\gamma\delta}+ {\rm e}^{i (\kvec + \Kvec)\cdot (\gamma-\delta)}{\rm e}^{i \kvec \cdot (l-m)}\langle c_{l\gamma\sigma}^\dagger c_{m\delta\sigma}^{\phantom{\dagger}}\rangle_0  Y_{\gamma\delta\sigma}\right]\right|^2}{\sum_{l\gamma}\left[n_{l\gamma\sigma}+\dfrac{1-X_{\gamma}}{X_{\gamma}}\right] + \sum_{l\gamma,m\delta}\left[\left({\rm e}^{i\kvec \cdot (l-m)}{\rm e}^{i(\kvec + \Kvec) \cdot (\gamma-\delta)} \langle c_{l\gamma\sigma}^\dagger
c_{m\delta\sigma}^{\phantom{\dagger}}\rangle_{0} - n^0_{l\gamma\sigma}\delta_{lm}\delta_{\gamma\delta}\right)Z_{\gamma\delta\sigma}\right]} \ , 
\label{MmYapp}
\end{eqnarray}
with
\begin{eqnarray*}
Y_{\gamma\delta\sigma} = \dfrac{\left[\sqrt{n_{l\gamma,-\sigma}}\sqrt{\dfrac{1-X_{\gamma}}{X_{\gamma}n_{l\gamma\sigma}}}+\sqrt{(1-n_{l\gamma})} \right]\sqrt{(1-n_{m\delta})}}{\sqrt{(1-n_{l\gamma\sigma})(1-n_{m\delta\sigma})}} \ . 
\label{Ylmapp}
\end{eqnarray*}
\end{widetext}

\section*{Appendix 2}\label{sec:appendixB}

\subsection{Monte Carlo results for projected superconductors}\label{MC:projectSC}

As aforementioned in Sect.~\ref{sec:Montecarlo}, we performed computations similar to the ones reported in Ref.~\onlinecite{Fuk05}, with the aim to understand better some topics discussed in that paper. 
For the superconductors, the lattice is not subdivided in two sublattices as for the case of the AFM, thus we drop the labels $\alpha$ and $\beta$ to indicate the lattice sites in the formalism of this Appendix.

We use the uncorrelated state $|\Psi_0\rangle$ obtained as the ground state of the Hamiltonian $H^0_{BCS}$:
\begin{eqnarray}
H^0_{BCS}&=&\sum_{ij\sigma}t_{ij}c^{\dagger}_{i\sigma}c^{\phantom{\dagger}}_{j\sigma}-\mu \sum_{i\sigma}c^{\dagger}_{i\sigma}c^{\phantom{\dagger}}_{i\sigma} \nonumber \\ &+&\sum_{ij}\Delta_{ij}(c^{\dagger}_{i\uparrow}c^{\dagger}_{j\downarrow}+\textrm{h.c.}) \ .
\label{HBCS}
\end{eqnarray}

We consider projected BCS-superconducting states on a square lattice, with the nearest-neighbour hopping $t_{ij}=-t$. Then, we define our partially projected wave functions as    
\begin{eqnarray}
|\Psi'_{l}\rangle &=& P'_{l} |\Psi_0\rangle \nonumber  \\
 P_{l}' &=& \prod_{i \ne l}(1-n_{i\uparrow} n_{i\downarrow}) \nonumber \\ 
|\Psi_0\rangle &=& \prod_{\kvec}[u_{\kvec} + v_{\kvec}c^{\dag}_{\kvec\uparrow}c^{\dag}_{-\kvec\downarrow}]|0\rangle \ , 
\label{BCS0}
\end{eqnarray} 
with 
\begin{eqnarray*}
 u^2_{\kvec} &=& \dfrac{1}{2}\Big(1+ \xi^0_{\kvec\sigma}/\sqrt{(\xi^0_{\kvec\sigma})^2+ \Delta^2_\kvec}\Big) \\   
v^2_{\kvec} &=& \dfrac{1}{2}\Big(1-\xi^0_{\kvec\sigma}/\sqrt{(\xi^0_{\kvec\sigma}))^2+ \Delta^2_\kvec}\Big) \ , 
\end{eqnarray*} 
and with $\xi^0_{\kvec\sigma}$ = $\epsilon^0_{\kvec\sigma} - \mu = -2t (\cos k_x + \cos k_y) - \mu$, being $\mu$ the chemical potential. In this Appendix the reservoir site is indicated with $l$. 
 
We check the accuracy of the GA for projected superconductors for two different cases.  
Upon applying a finite $\Delta$ [see Eq.~(\ref{HBCS})], we obtain BCS superconducting states and in particular we investigate what happens for the case of s-wave and d-wave symmetry of the gap.

As we did in Sect.~V for the projected AFM states, we consider the behaviour of the double occupancy $d_l$ of the reservoir site in VMC in order to test the limits and the reliability of the GA. 

\begin{center}
\begin{figure}[tp]
\includegraphics[width=8.65cm]{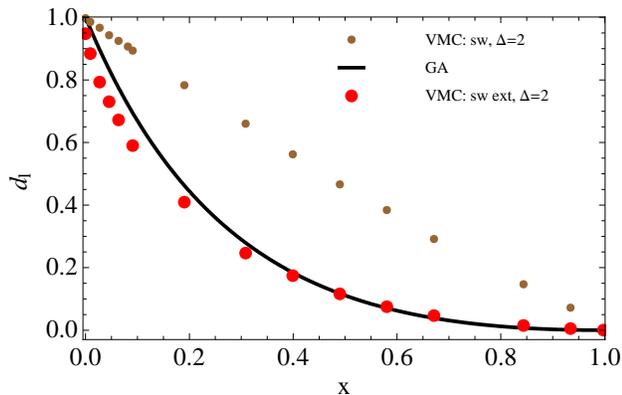}   
\caption{Double occupancy $d_l$ of the unprojected reservoir site as a function of the hole doping $x$ for the projected s-wave superconductors on a 882-site lattice: GA and VMC results are compared for the field $\Delta = 2$ in the case of onsite (sw) and nearest-neighbour BCS interaction (sw ext).} 
\label{dSC-sw}
\end{figure}
\end{center}

{\bf s-wave BCS superconductors.} In this case the superconducting gap is isotropic.
In Fig.~\ref{dSC-sw}, we show the corresponding behaviour of the double occupancy of the reservoir site upon increasing the doping. In particular, we consider two different types of s-waves: the one given by onsite interaction $\Delta$ (recovering the results of Ref.~\onlinecite{Fuk05}) and the extended s-wave given by nearest-neighbour BCS interaction.   

The trends of the double occupancy for the two types of s-waves are very different: for the extended s-wave the VMC results are very well approximated by the analytical ones for all the dopings. On the contrary, for the local s-wave the double occupancy is highly enhanced with respect to the GA and for this specifical value of $\Delta$ the curve is quasilinear with the convexity tendentially downwards.     

These results imply that the GA fails completely as a good method to represent the behaviour of the projected onsite s-wave superconductor for any doping and this was quite unexpected, in that the GA should work well for the wave functions which do not exhibit spatial inhomogeneities and should be extremely accurate in the diluted limit, while this is not the case. Nevertheless, we can explain the physics responsible for this peculiar trend not captured within the GA. In fact, the onsite BCS interaction favours the onset of double occupancies with respect to the case of the intersite interaction which places the electrons preferentially on distinct sites. This scenario is clear for uncorrelated wave functions but actually it can be exported also in presence of our Gutzwiller projector; the different behaviour between the VMC trend for the pure s-wave and the trends for the extended one (in VMC and in GA) becomes robust upon increasing the size of $\Delta$. In the limit of infinite applied field $\Delta$, the double occupancy of the reservoir site 
for the local s-wave will be  maximal. In particular, for the uncorrelated local s-wave all the sites are unprojected and exhibit the double occupancy $d=n/2$; in the  strongly correlated case $d_l = 1$ for $\Delta = \infty$. The GA is not sensitive to this limit for the projected local s-wave, as we can see in Fig.~\ref{dSC-sw}: the quality of the approximation is highly unsatisfactory already for a finite value of $\Delta$, quite far from the asymptotical regime of $\Delta = \infty$.    


   
\begin{center}
\begin{figure}[tp]
\includegraphics[width=8.65cm]{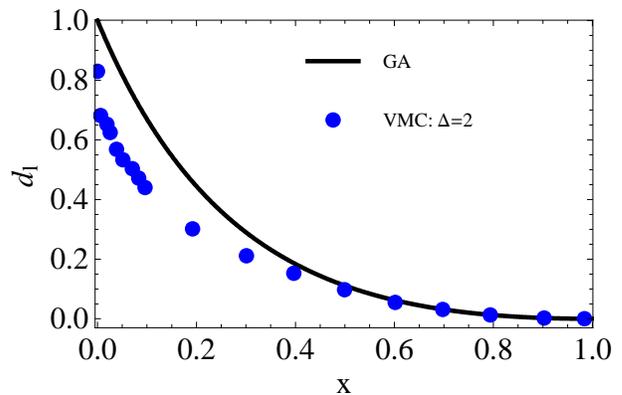}
\caption{Double occupancy $d_l$ of the unprojected reservoir site as a function of the hole doping $x$ for the projected d-wave superconductors on a 2500-site lattice: GA and VMC results are compared for the field $\Delta = 2$.} 
\label{dSC-dw}
\end{figure}
\end{center} 

{\bf d-wave BCS superconductors.} The superconducting gap shows this symmetry in the momentum space: $\Delta_\kvec = \Delta (\cos k_x - \cos k_y)$.

In Fig.~\ref{dSC-dw}, we report the corresponding behaviour of the double occupancy $d_{l}$ upon increasing the doping. For this case, the results presented in Ref.~\onlinecite{Fuk05} were not univocal, in the sense that the quantity $a_{k=0} = u_{k=0}/v_{k=0}$ is indeterminate and thus the authors show the behaviour of the double occupancy for two possible values of $a_{k=0}$: 0 and {\em large}, {\em i.e.} larger than all the other coefficients $a_{k}$ (see their Fig.~3). In general, they presented all their other results for $a_{k=0}$ large [as in their Fig.~2]. Actually, their choice was an ansatz and the quantitative discrepancy in the two different cases were quite relevant. In our computations, we obtain results compatible with theirs with larger $a_{k=0}$ and this supports their choice. 

Then one more open issue was left in the former VMC investigation by Fukushima {\em et al.}: in fact they noticed that the trend of the double occupancy close to half-filling was extremely more sensitive to the size effects if compared to what occurs for lower fillings. This means that we are to perform computations raising the number of the sites $L$ to several thousands in order to have a meaningful comparison with the GA, which is valid in the thermodynamical limit. According to the GA, $d_{l}$ has to be 1 for $n=1$: in Ref.~\onlinecite{Fuk05} they found $d_{\gamma}$ between 0.6 and 0.8 for $L$ between 1000 and 2000 sites. A further increase in the size is practically very difficult because numerical computations become long-time demanding, and a too limited statistics can cause not ergodic results. For $L=2500$, $d_{l}$ overcomes 0.8. According to the size-scaling trend, we are quite confident that this quantity is slowing tending to 1, but we can not claim this as a sure finding. This is the only case where we have a result so dependent in size also for very large lattices: typically the results for $L>500$ are quantitatively very similar to the ones for other larger $L$. Clearly, one would be highly interested to fix this issue with VMC computations: a thermodynamical value of $d_{l} \neq 1$ for the projected BCS states in d-wave at half-filling would imply that the GA result is not realized in this case, while it is confirmed by the VMC computations for the other projected wave functions tested so far at half-filling.

\end{document}